# Percolation properties in a traffic model

Feilong Wang[1,2], Daqing Li[1,2 (a)], Xiaoyun Xu[1,2], Ruoqian Wu[1,2] and Shlomo Havlin[3]

[1] *School of Reliability and Systems Engineering, Beihang University, Beijing 100191, China*
[2] *Science and Technology on Reliability and Environmental Engineering Laboratory, Beijing 100191, China*
[3] *Department of Physics, Bar Ilan University, Ramat Gan 52900, Israel*

**Abstract** – As a dynamical complex system, traffic is characterized by a transition from free flow to congestions, which is mostly studied in highways. However, despite its importance in developing congestion mitigation strategies, the understanding of this common traffic phenomenon in a city-scale is still missing. An open question is how the traffic in the network collapses from a global efficient traffic to isolated local flows in small clusters, *i.e.* the question of traffic percolation. Here we study the traffic percolation properties on a lattice by simulation of an agent-based model for traffic. A critical traffic volume in this model distinguishes the free-state from congested state of traffic. Our results show that the threshold of traffic percolation decreases with increasing traffic volume and reaches a minimum value at the critical traffic volume. We show that this minimal threshold is the result of longest spatial correlation between traffic flows at the critical traffic volume. These findings may help to develop congestion mitigation strategies in a network view.

The urban transportation system, as one of the critical infrastructures, to a great extent, facilitates the performance in many aspects of human activities. However, traffic transition, from free flow to seriously congested state [1], occurs frequently in our daily life, deteriorates the system's efficiency, and even causes related problems such as pollution and waste of time [2]. To date, various models based on different approaches have been proposed to simulate the traffic flow and to achieve a comprehensive picture of the traffic dynamics including the features of traffic transition. These modelling methods [1, 3-8] mainly focus on two scales: macroscopic and microscopic. Macroscopic studies have been dedicated to reveal the traffic dynamics through the theory of kinetic gas or fluid dynamic; on the other hand, the microscopic study aims at capturing behaviors of the individual cars and simulating the basic interactions among them. Based on these methods, phenomena of jamming and its relevance to systematic deterioration of traffic are extensively studied.

However, the mechanisms of how a network-scale global traffic disintegrates into local flows are still unclear. Indeed, the global traffic system may be improperly improved due to the lack of knowledge on the interactions of local congestions and their influence on the macroscopic scale. We propose here to study the organization of global traffic flow on a lattice model using percolation approach [9, 10]. Percolation theory is a useful tool for studying the network organization and disintegration, with extensive applications as diverse as diffusion, epidemics, and system robustness. In the percolation theory, the threshold $p_c$ is defined as the critical point for the phase transition from network connectivity to fragmentation. This threshold can be calculated theoretically based on analytical considerations, or by simulations. Recently, by constructing a traffic dynamical network, it is found that the organization of city traffic can be considered as a percolation-like transition [11]. This result is based on analysis of city traffic real data and it was revealed that the city-scale global traffic is dynamically composed of different local traffic clusters, which evolve during the day. However, the relation between traffic parameters (volume and routing choice) and the system percolation properties has not been studied systematically.

In the present study, based on an agent-based model developed by Echenique *et al*. [12] and extensive simulations, we study the relation between percolation organization and traffic by varying two main traffic factors: the traffic volume *OD* and the path-routing choice. It has been found [12] that with the increase of traffic volume *OD* in the network, the traffic will undergo a phase transition between free state and congested state at a critical value $OD_c$. We introduce here a percolation parameter *p,* which characterizes the efficiency of the global traffic (for the detailed definition of *p*, see the section Traffic percolation below). We analyze here the critical threshold of traffic percolation $p_c$ as a function of *OD* and find that $p_c$ has a sharp minimum value at the traffic transition threshold $OD_c$, indicating the existence of the most efficient organization of global traffic flow. The value of $p_c$ is influenced by the routing choice. We reveal here that this behavior of $p_c$ results from the spatial correlation of traffic load, which is maximal at traffic transition $OD_c$. We also analyzed the distribution of node betweenness in the giant component at percolation criticality. Finally, we discuss how our findings may provide insights to the practical management of urban traffic.

**Traffic model.** – Here, we focus on simulations of a traffic dynamical model on a diluted periodical square lattice. In the square lattice, links and nodes represent road sections and their intersections respectively. To take into account the disorder factors due to traffic control measures and diverse road conditions, a certain fraction of randomly selected links are initially removed to avoid a purely regular network and weights are added to the remaining links. Specifically, in our study we randomly removed 10 percent of edges. Link weights follow a Gaussian distribution $G(\mu, \sigma)$, where average $\mu$ is 1 and standard deviation $\sigma$ is 0.33 in the presented study. Other values of $\sigma$ can yield similar results.

In our study, for simulating the adaptive traffic collective behaviours, we use an agent-based model [12]. To simulate the traffic dynamics in the model, at each time step a given number ($OD$) of packets (cars) are created and delivered in the network following a given routing strategy. Both the origin and destination of each packet are selected randomly from all the nodes. The size of packet is chosen randomly following a Gaussian distribution. For simplicity, the parameters of the Gaussian here are the same as the distribution of link weight. Each packet contributes to the load of its current node by the addition of its size. It is assumed that each node has an identical capability to deliver only one packet at each time step, with the first-in-first-out basis. At each time step, the delivery process of packets on the network is carried out in a parallel update algorithm. After the delivery of all packets and the update of load configuration, we start a new iteration and the time step $t$ is increased by one. As the consideration of drivers' capability to circumvent congested areas, a packet, setting off from its current site, selects its travel path dynamically according to a traffic-aware routing scheme discussed below and will be removed after reaching its destination.

The traffic-aware routing strategy [12] incorporates the local traffic awareness with global information in the delivery algorithm, representing drivers' behavior on path selection according to the dynamical traffic situation. When a packet with destination $j$ staying on node $i$ is choosing its next stop among the $i$'s neighbors, an effective distance $e_{kj}$ between a neighbor $k$ of $i$ and $j$ is introduced and defined as

$$e_{kj} = hd_{kj} + (1-h)l_k \quad (1)$$

Here, $d_{kj}$ is the optimal distance, defined here as the length of path minimizing the sum of link weights between $k$ and $j$, which is computed using Dijkstra algorithm [13, 14]. And $l_k$ is the current load of node $k$, which is the total contribution made by the size of all packets accumulated as a queue on the node $k$. $h$ is a tunable parameter between 0 and 1, that accounts for the degree of traffic awareness incorporated in the delivery algorithm. Following this definition, the neighbor $k$ with minimal effective distance $e_{kj}$ will be selected as the packet's next stop. In our simulation, if two or more paths have identical effective distance, we randomly select one of these paths. The routing scheme in Eq. (1) gives us a chance to capture the path-selection behavior of the drivers, which considers both the global knowledge of network's topology and the current traffic situation ahead. Notice that a small $h$ means that drivers care more about the local traffic situation and are willing to circumvent the congested districts.

For each simulation of traffic flow in our study, we run the model for sufficient time steps to reach a stationary state, which is indicated by the system's order parameter, $\rho$, defined as [12],

$$\rho = \lim_{t \to \infty} \frac{L_N(t+\Delta_t) - L_N(t)}{\Delta_t OD} \quad (2)$$

Where $L_N(t)$ is the total load in the network at time step $t$ and $\Delta t$ is the observation time window. The infinite limit of $t$ ensures that we reach a stationary state, where $\rho$ becomes stable with time. As an indicator of traffic situation of the network, $\rho$ estimates the balance of the outflow and inflow. The case $\rho = 1$ represents a completely congested traffic, since no car can reach its destination, and $\rho = 0$ refers to a free-flow state.

In the following, we focus on the features of traffic percolation as a function of traffic volume $OD$ and the routing strategy $h$.

**Traffic percolation.** – To understand and reveal how the city-scale traffic flow breaks down into clusters of local flows, we apply a percolation analysis on the model simulation results. At the stationary state of each traffic flow simulation, we record the load on every node. Then a threshold, $m$, which varies from the minimum to the maximum of the node load, is defined to act as the only control parameter in our percolation analysis. The state $s_i$ of the node with load $l_i$ will be grouped into one of two classes: free-state for $l_i \leq m$ or congested for $l_i > m$, i.e.

$$s_i = \begin{cases} 1, & l_i \leq m \\ 0, & l_i > m \end{cases} \quad (3)$$

The fraction of free-state nodes ($s_i = 1$) in the network, which can be considered as the occupation fraction $p$ in percolation, increases as we raise the threshold $m$ gradually. In this process, we choose an appropriate interval ($\Delta m = 0.002$ in our study) so that the increase of $m$ is small enough to keep at most one node's state modified (one $s_i$ changes from 0 to 1). For nodes with equal load, we randomly select one of them and update its state. As we increase the threshold $m$, clusters of free-state nodes ($s_i = 1$) will emerge and we can observe the occurrence of percolation process in the network: for small $m$, almost all nodes are considered as congested and only small clusters will appear; for large $m$, small clusters will merge into larger clusters, showing the organization process of local traffic flow. The emergence of the giant cluster indicates the occurrence of percolation transition and the formation of global traffic flow. At a certain fraction of free-state nodes $p_c$ (determined by $m_c$), the second-largest cluster reaches its maximum, which signifies the formation of global traffic flow [9] (fig. 1a and 1c). According to percolation theory, this value $p_c$ refers to the critical threshold of traffic percolation [9]. This clustering process is illustrated in fig. 1b and 1d.

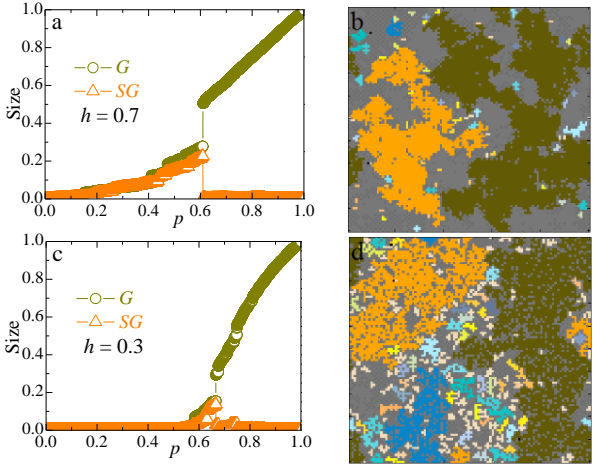

Fig. 1: Percolation of traffic flow on the lattice. In (a) and (c), we show the size of both the largest functional cluster $G$ and the second-largest cluster $SG$ changes with $p$ for different $h$: (a) $h = 0.7$ and (c) $h = 0.3$. Note that we define here $G$ as the size of largest cluster among all clusters in the diluted lattice obtained for a given $m$. In (b) and (d), we show the percolation clusters of the network near $p_c$ for different $h$: (b) $h = 0.7$ and (d) $h = 0.3$. The largest cluster is marked in olive green and the second largest one is marked in orange. All the grey nodes are those congested and removed from the network. Other colors refer to other small functional clusters. The results here are obtained on a lattice (100×100) at $OD = 150$.

In a real urban transportation system, the dynamical traffic volume is a parameter of great importance, because of its direct effect on the emergence of congestions [1, 15-17]. To observe the effect of traffic volume on traffic percolation, we first analyze the order parameter $\rho$, Eq. (2), of the system at the stationary state as a function of traffic volume, $OD$. We can observe a transition at $OD_c$ from free traffic to congestion as $OD$ increases (fig. 2a and 2b), similar to that found in [12]. A critical traffic volume $OD_c$ can be found, above which systematic congestions occur. We next study how these traffic parameters affect the traffic percolation properties.

In the analysis of the traffic percolation properties, $p_c$ can be regarded as a quantitative measure for the efficiency of the global traffic flow [11]. In the present study, we explore the influence of traffic volume on $p_c$ for different $OD$ (fig. 2a and 2b). Strikingly, our findings suggest that $p_c$ initially decreases non-linearly with increasing $OD$. A sharp minimum $p_c$ is clearly observed at the critical $OD$, indicating that the global flow can be efficiently maintained with minimal number of local flows at criticality. We demonstrate that the results are similar under different path-selecting behaviors represented by $h$ (fig. 2b), with different value of minimal $p_c$. Note also, that only near $OD_c$ the deviation from random percolation is significant and reaches the maximal value at $OD_c$. When we recover the classical random percolation by shuffling the node loads, the interesting non-monotonic behavior and the sharp minimum of $p_c$ disappear. We next study the origin of this striking minimum of $p_c$ in the traffic organization.

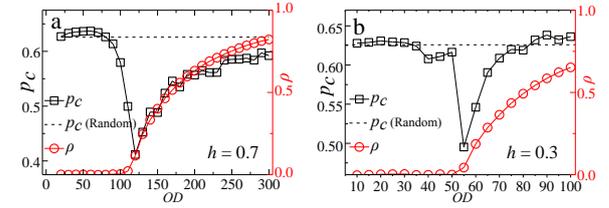

Fig. 2: The critical threshold $p_c$ and order parameter $\rho$ as a function of $OD$ for different $h$ (0.7 and 0.3 in (a) and (b) respectively). The critical $OD$, $OD_c$, are near 120 and 55 for $h = 0.7$ and 0.3 respectively. For comparison, we show the classical random percolation (in dash line) by shuffling the node loads. Simulations have been carried out on a lattice of size 100×100 with averages over 200 realizations. We choose our lattice with linear size $L = 100$, because bigger system sizes are very time consuming.

**Correlation length in traffic.** – Correlations have been found between traffic flow in highways [1]. To understand how the traffic volume influences $p_c$, in particular the origin of strong deviation from random percolation near $OD_c$, we investigate the spatial correlation length [18-21] of the traffic load. Different from the random percolation, the local flows interact with correlations. It is interesting to study how the correlation of traffic flows influences the threshold $p_c$ of traffic percolation. We calculate the spatial correlation of loads with different $OD$ in the network. The correlation function, $C(r)$, measures the correlation between traffic loads at distance $r$ [19-21]:

$$C(r) = \frac{1}{\sigma^2} \frac{\sum_{ij,\, i \in S}(x_i - \bar{x})(x_j - \bar{x})\delta(r_{ij} - r)}{\sum_{ij,\, i \in S}\delta(r_{ij} - r)} \quad (4)$$

Here, $x_i$ is the load on node $i$. $\bar{x}$ is the average load over all nodes and $\sigma^2 = \sum_{i,\, i \in S}(x_i - \bar{x})^2 / N_S$ is the variance. Here, $S$ is the set of all nodes in the network of size $N_S$. $r_{ij}$ is the Euclidean distance between nodes $i$ and $j$. The delta function selects nodes at distance $r$. Following this definition, positive values of $C(r)$ mean that traffic situations at distance $r$ are positive correlated, and vice versa.

We show in fig. 3a and 3b that the correlation decays in different way for different traffic volume. The correlation length $\xi_l$, is defined as the distance, where $C(r)$ becomes zero for the first time [19], representing the scale of the correlated domains (fig. 3a and 3b). We show the correlation length as a function of *OD* values in fig. 3c and 3d, and find that the correlation length increases when critical traffic volume, $OD_c$, is approached. The correlation length of traffic load reaches its maximum at $OD_c$ for different $h$, suggesting the traffic flow is correlated with the longest range at the critical point.

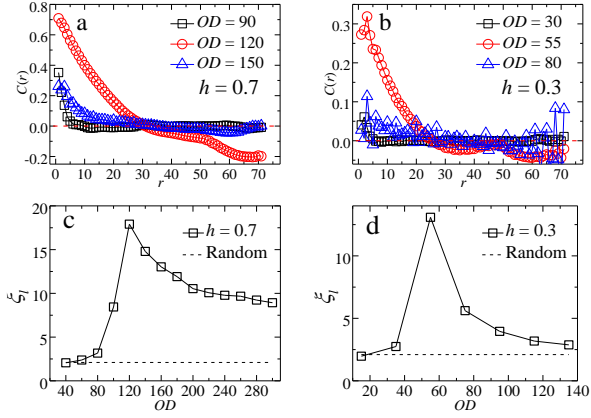

Fig. 3: The spatial correlation of traffic load. In (a) and (b), we show the spatial correlation of loads for different *OD*. In (c) and (d), we show the correlation length of loads $\xi_l$, as the function of traffic volume *OD* for different values of $h$. Note that $\xi_l$ reaches its maximum at the critical *OD*. Simulations have been carried out on a lattice of size $100 \times 100$ with 100 realizations averaged.

When correlation is added artificially in a percolation model [22, 23], some percolation properties, at the presence of long-range correlation, are found different from the uncorrelated case, including the percolation threshold $p_c$. In our study, when correlation of traffic percolation appears as a result of self-organized flow interaction, it is also observed that the percolation threshold $p_c$ is significantly decreased at the presence of longest correlation length when $OD_c$ is approached. Thus, the results of load correlation here offer an explanation for the microscopic origin of the minimum of $p_c$ at the critical traffic volume $OD_c$.

**Optimal paths in the giant component of traffic percolation. –** To get further insight on how our findings can benefit the management of urban traffic, we explore the distribution of optimal paths in the network. The betweenness centrality $C$ is a metric to characterize the importance of a node in terms of transportation, which can be determined by counting how many times it is used by all the optimal paths [24]. The optimal path between two nodes is the path with the minimal cost among all available paths. The calculation of optimal path here considers both the static link weight and dynamical node weight that is the total size of packets on this node, in order to have a comprehensive consideration of both the network topology and the current traffic situations.

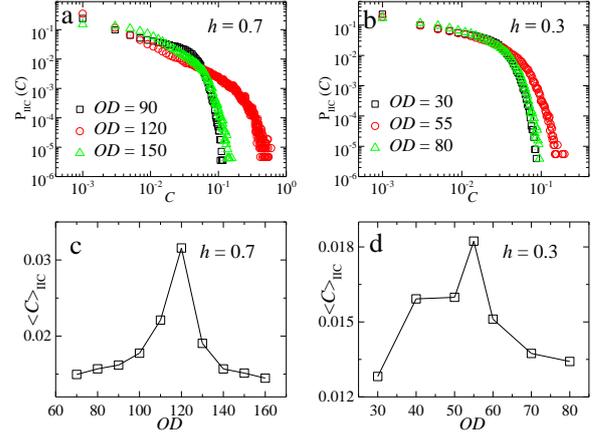

Fig. 4: The distribution of the betweenness centrality in the giant component. (a) and (b) demonstrate the distribution of betweenness, $P_{IIC}(C)$, for different *OD*. In (c) and (d), we show the averaged betweenness of nodes in the giant component $<C>_{IIC}$ as a function of *OD*. Simulations have been carried out on a lattice of size $100 \times 100$ with 50 realizations averaged.

In order to provide suggestion on how the percolation properties found here may improve real traffic, we now focus on the study of nodes betweenness $C$ on the giant percolation cluster (IIC), which behaves as the backbone of the network [25]. Comparing to the traditional calculation of the optimal path [26, 27], which mainly considers the static weight of links, here the calculation also includes the dynamical load of packets on each node as node weight, which changes with time and load condition. We calculate the probability density function of $C$ for nodes in the giant component ($P_{IIC}(C)$) at $p_c$ for a given *OD*, which seems consistent with results in [25]. Results also show that $P_{IIC}(C)$ varies significantly for different *OD*. For $h = 0.7$, we find that $P_{IIC}(C)$ decays quite slow at the critical value $OD_c$, while $P_{IIC}(C)$ for *OD* deviating from $OD_c$ decays faster. We get similar results for $h = 0.3$, where $P_{IIC}(C)$ at $OD_c$ decay faster than the case with larger $h = 0.7$.

To further illustrate the influence of *OD* on the distribution of optimal paths, we calculate $<C>_{IIC}$, the average $C$ over all nodes in the giant component at $p_c$ for a given *OD*. $<C>_{IIC}$ measures how many optimal paths on average pass through the giant percolation component. We find that $<C>_{IIC}$ changes significantly with *OD*, with a sharp maximum at $OD_c$ (fig. 4c and 4d). The relatively high centrality of nodes in the giant component at $OD_c$ confirms the result of minimal percolation threshold $p_c$, both of which suggest the highest efficiency at critical volume. The reason is that the spatial correlation of traffic load $\xi_l$ (fig. 3) has the maximal value at $OD_c$, leading to the accumulation of high betweenness nodes in the giant component. Given the awareness to the traffic situations, drivers are able to reroute their travel paths, to take advantage of free-flow regions provided by the high betweenness in the giant component.

In conclusion, by capturing some realistic behaviors of drivers in the daily traffic using an agent-based model featured with a traffic-aware routing strategy, we investigate the properties of traffic percolation under the influence of two traffic factors: the traffic volume and the drivers' path-routing choice. Interestingly, we find that the maximal efficiency of traffic occurs at $OD_c$, where traffic percolation has a minimal threshold $p_c$. We show that this results from the enhanced strength of spatial correlation of loads at criticality. At the presence of the relationship between percolation and traffic, we reveal that the high accumulation of optimal paths in a giant percolation component provides a possible routing choice to avoid congestions, which may suggest ways of mitigating the traffic congestions. The particular definition of effective distances [12] is essential to capture the realistic behaviors of drivers in the daily traffic. Given that the linear dependence of routing strategy in Eq. (1) is assumed, a study is needed in the future to find out the dependence function and its effect on traffic percolation from real traffic data. Furthermore, inspired by the successful application of network-view approaches to the study of realistic complex systems [28-30], we expect the future exploration on the comparison of our findings with the analysis from real data.

\*\*\*

This work is supported by Collaborative Innovation Center for industrial Cyber-Physical System and the National Basic Research Program of China (2012CB725404). D.L. acknowledges support from the National Natural Science Foundation of China (Grant 61104144) and the Fundamental Research Funds for the Central Universities. S.H. thanks Defense Threat Reduction Agency (DTRA), Office of Naval Research (ONR) (N62909-14-1-N019), United States-Israel Binational Science Foundation, the LINC (Grant 289447) and the Multiplex (Grant 317532) European projects and the Israel Science Foundation for support.